\documentclass[aps,twocolumn,superscriptaddress,nofootinbib,floatfix]{revtex4-2}

\usepackage{amsmath}
\usepackage{mathrsfs}
\usepackage{amssymb}
\usepackage{bm} 
\usepackage{graphicx,grffile,textcomp}
\usepackage{dcolumn} 
\usepackage[usenames,dvipsnames]{xcolor}
\usepackage{multirow}
\usepackage{epstopdf}
\usepackage{mathtools} 
\usepackage{tabularx}
\usepackage{soul} 
\usepackage{microtype}
\usepackage{comment}
\usepackage[version=4]{mhchem} 
\usepackage{array}
\usepackage{float}
\usepackage{wrapfig} 
\usepackage{url}
\usepackage{hyperref}
\usepackage{cleveref}
\usepackage[utf8]{inputenc}
\usepackage[T1]{fontenc}
\usepackage{bbold}

\hypersetup{colorlinks=true, linkcolor=blue, urlcolor=blue, citecolor=blue}

\definecolor{red-1}{rgb}{0.92, 0.035, 0.36}

\renewcommand{\Re}{\operatorname{Re}}
\renewcommand{\Im}{\operatorname{Im}}

\newcommand{\vect}[1]{\mathbf{#1}}

\begin{document}

\title{Charge separation at liquid interfaces}

\author{Arghya Majee}
\affiliation{Max Planck Institute for the Physics of Complex Systems, 01187 Dresden, Germany}

\author{Christoph A. Weber}
\email{corresponding authors: christoph.weber@physik.uni-augsburg.de and julicher@pks.mpg.de}
\affiliation{Faculty of Mathematics, Natural Sciences, and Materials Engineering: Institute of Physics, University of Augsburg, Universit\"atsstr. 1, 86159 Augsburg, Germany}

\author{Frank J\"ulicher}
\email{corresponding authors: christoph.weber@physik.uni-augsburg.de and julicher@pks.mpg.de}
\affiliation{Max Planck Institute for the Physics of Complex Systems, 01187 Dresden, Germany}
\affiliation{Center for Systems Biology Dresden, 01307 Dresden, Germany}
\affiliation{Cluster of Excellence Physics of Life, TU Dresden, 01062 Dresden, Germany}

\date{August 5, 2024}

\begin{abstract}
We present a theory for phase-separated liquid coacervates with salt, taking into account spatial heterogeneities and interfacial profiles. We find that charged layers of alternating sign can form around the interface while the bulk phases remain approximately charge-neutral. We show that the salt concentration regulates the number of layers and the amplitude of the layer's charge density and electrostatic potential. Such charged layers can either repel or attract single-charged molecules diffusing across the interface. Our theory could be relevant for artificial systems and biomolecular condensates in cells. Our work suggests that interfaces of biomolecular condensates could mediate charge-specific transport similar to membrane-bound compartments.
\end{abstract}

\maketitle

\section{Introduction\label{Sec:1}}

Spatial organization of macromolecules is essential for regulating cellular processes. Most biochemical processes involve charged macromolecules like  DNA, RNA, and proteins, that are immersed in a multi-component, aqueous electrolyte solution containing various salt ions. These biomolecules assemble into specific compartments that provide unique environments to perform certain tasks \cite{Hyman2014, Shin2017}. Moreover, compartmentalization of prebiotic components and chemical reactions was also proposed as a selection mechanism at the origin of life \cite{Booij1956, Oparin1957, Szostak2001, Bartolucci2023}.

Compartments in cells can either be membrane-bound, or membrane-less liquid droplets that can disperse and reform for varying conditions in cells \cite{Brangwynne2009, Guillen2020, Fritsch2021}. Such intra-cellular droplets are called biomolecular condensates and often form via liquid-liquid phase separation \cite{Hyman2014, Shin2017, Banani2017, Franzmann2019}. Many biomolecular condensates in cells contain negatively charged RNA and positively charged proteins \cite{Franzmann2018} suggesting coacervation as a mechanism underlying their formation \cite{Lu2020}. Indeed, in vitro experiments show that condensate properties are strongly affected by salt concentration \cite{Jawerth2018,Hubatsch2021, Mccall2023}. Coacervates are liquid droplets composed of charged macromolecules and counterions interacting via electrostatic interactions. In physical chemistry, they are classified into two categories \cite{Abbas2021}: \textit{simple} coacervates formed by a single type of macroion and its counterions, and \textit{complex} coacervates due to electrostatic interactions between oppositely charged macroions. Coacervation as a mechanism to spatially organize and select molecules was already proposed at the beginning of the twentieth 
century as an organizing principle for the molecular origin of life~\cite{Booij1956, Oparin1957}.

\begin{figure}[b]
   \includegraphics[width=8.0cm]{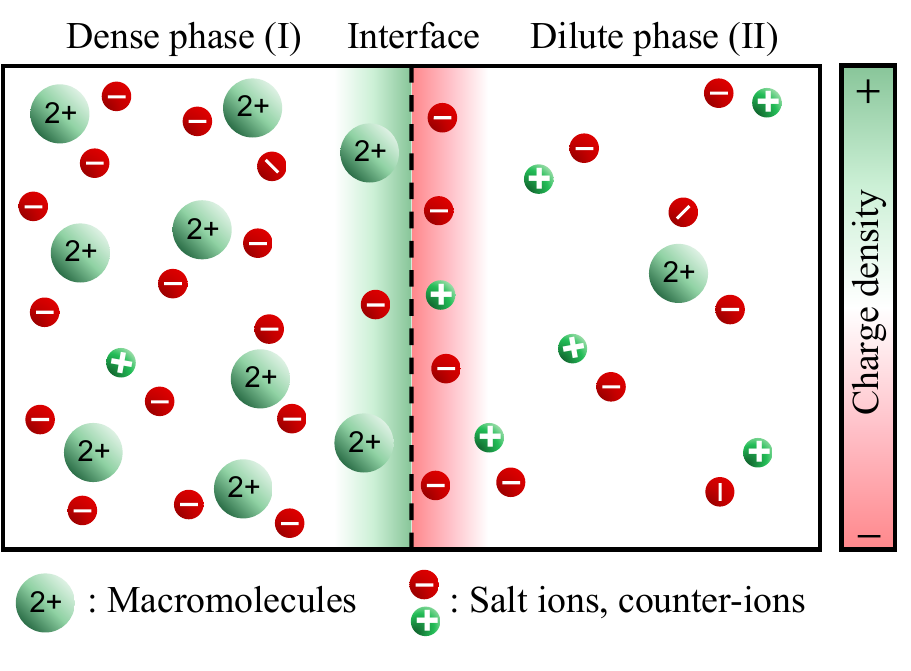}
   \caption{\textbf{Schematics of the considered phase-separated electrolyte mixtures.} Macromolecule-rich phases (phase I) can form in an initially homogeneous saline solution with macromolecules (big green circles). Macromolecules are depicted to be positively charged with counter ions being anions. Dissociation of salts also generate co/counter-ions. Charges can separate with a charge density that deviates from electro-neutrality, in particular around the interface (indicated by the color gradients).}
   \label{Fig:1}
\end{figure}

A first theoretical model for coacervates was proposed by Voorn-Overbeek \cite{Overbeek1957}. Within this model, phase separation results from a competition between the mixing entropy and the electrostatic interactions between the oppositely charged molecules, which is described using the Debye-H\"uckel approximation in the Poisson-Boltzmann equation treating ions as dilute components. However, when coacervates form, charged macromolecules and salt ions are in general non-dilute and interactions among all charged components are essential, in particular at physiological conditions. Interactions were shown to indeed affect the distribution of macroions close to charged surfaces \cite{Bostroem2001, Adar2017}. Most theories on electrolytes study flat and colloidal charged surfaces that are solid \cite{Dean2014, Markovich2021}. In contrast, coacervates provide phase boundaries separating liquid phases that are soft and permeable.  

Recent experimental studies show that coacervate phase boundaries give rise to rich physical behavior \cite{Spruijt2009, Banerjee2017, Jawerth2018, Alshareedah2021, Welsh2022, Agrawal2022}. Condensates can carry a surface charge and thus move in an applied electric field due to electrophoresis \cite{Banerjee2017, Welsh2022, Agrawal2022, Haren2024}. The surface charge and the associated zeta potential were also recognized as one of the factors determining droplet stability \cite{Welsh2022}. Not only that, the interfacial tension of protein droplets depends on salinity as well as on the Donnan potential difference that exists between two coexisting phases \cite{Vis2015, Jawerth2018}. Such studies indeed highlight the importance of electrostatics and interfaces in phase-separated, coacervate systems. Recently, Zhang and Wang have proposed a theory that can explain the origin of interfacial charges in asymmetric complex coacervates \cite{Zhang2021}. This work closely parallels the concepts provided by Onuki \cite{Onuki2006} for partitioning of diluted salt with asymmetries in Born solvation energies across phases in binary mixtures.

In this work, we present a theoretical framework similar to Refs.~\cite{Onuki2006, Zhang2021, Aerov2007-1} and apply it to non-dilute mixtures composed of charged macromolecules and salt ions capable of forming coexisting phases, i.e., a coacervate phase-separated from a dilute phase. We account for the mean-field electrostatic interactions among all charged components and study the concentration profiles of components in each phase and in the interfacial domain. Our key finding is that charge separation occurs at interface while each phase is approximately charge-neutral (see an illustration in Fig.~\ref{Fig:1}) both for symmetric and asymmetric complex coacervates as well as for simple coacervates. The corresponding electrostatic potential shows a non-monotonic behavior with attractive wells and repulsive barriers. Strikingly, the electrostatic potential can also give rise to multiple layers of alternating charges. This implies that charged molecules may exhibit a complex transition kinetics through coacervate interfaces. We also present an analytical calculation to investigate the charge oscillations in the bulk phases and identify competing length scales of different physical origins that lead to such oscillatory profiles.

\section{Electrothermodynamics of phase separation\label{Sec:2}}

We consider an electrolyte mixture composed of charged macromolecules and counter-ions that can phase separate into a macromolecule-rich and a macromolecule-poor phase; see Fig.~\ref{Fig:1} for an illustration. Each  component, $i=1,...,M$, can carry a molecular charge $q_i e$ with $e$ denoting the positive elementary charge. Here we consider constant charge for each component and for simplicity do not account for chemical processes leading to charge regulation \cite{Arana2020, Majee2020, Celora2023}. Moreover, the electrolyte mixture is considered to be incompressible corresponding to constant molecular volumes $\nu_i$. Incompressibility implies that the condition $\sum_{i=1}^M\nu_i n_i(\vect{x}) = 1$ holds locally, where $n_i(\vect{x})$ denotes the concentration field of component $i$ at position $\vect{x}$. At constant temperature and pressure, this condition reduces the thermodynamic description to $(M-1)$ independent concentration fields  $n_i(\vect{x})$. 

All the components in the mixture collectively generate a dielectric medium with permittivity $\varepsilon=\varepsilon_r\varepsilon_0$, where $\varepsilon_0$ is the vacuum permittivity and $\varepsilon_r$ denotes the relative permittivity which in general depends on composition $\{n_i \}$. The distributions of charges give rise to an electrostatic potential $\psi(\vect{x})$, which is related to the concentrations $n_i(\vect{x})$ via the Poisson equation:
\begin{subequations}
\label{eq:theory}
\begin{align}
\nabla\cdot\left(\varepsilon \, \nabla\psi\right) =-\rho \, ,
\label{eq:1}    
\end{align}
where $\rho(\vect{x})=\sum_i q_i e n_i(\vect{x})$ is the local charge density. The presence of the charges in the system also naturally give rise to a length scale called the Bjerrum length $\ell_B=e^2/\left(4\pi\varepsilon k_BT\right)$ specifying the separation at which the Coulombic interaction between two elementary charges becomes comparable to the thermal energy $k_BT$ \cite{Adar2017}.

The electrothermodynamics of the mixture is governed by the free energy functional
\begin{align}
    &F\left[n_i, \psi\right] = \int d^3 x \left[ f\left(n_i\right)
    + \sum_i \frac{\kappa_i}{2} \left( \nabla n_i \right)^2
    + \frac{\varepsilon}{2} (\nabla \psi)^2 \right.\nonumber\\
    &\quad  \left. 
    + \lambda \bigg(  \nabla\cdot\left(\varepsilon \, \nabla\psi\right) +\rho \bigg)
    +\sum_i \mu_i^\text{el} \left(\frac{N_i}{V} - n_i\right)\right] \, ,
\label{eq:2}
\end{align} 
where $f$ is the free energy density. Moreover, $\kappa_i$  is related to the interfacial tension characterizing the energy contribution due to the spatial gradients of the components in the system \cite{Cahn1958, Safran2018, Weber2019}. We refer to $\kappa_i$ as the gradient cost for component $i$ in the following. We have neglected cross terms of the form $\nabla n_i\cdot\nabla n_j$ for simplicity. The term proportional to $\left(\nabla\psi\right)^2$ describes the electrostatic energy density arising from the charged components in the system. Moreover, $\lambda$ is the Lagrange multiplier to impose that the Poisson Eq.~\eqref{eq:1} is satisfied and $\mu_i^\text{el}$ is the Lagrange multiplier fixing the total number of particles $N_i=\int_V d^3x\,n_i(\vect{x})$ for each component $i$ with $V$ denoting the volume of the system. 

To describe interactions among all components we choose the Flory-Huggins (FH) free energy density
\begin{align}
   \frac{f\left(n_i\right)}{k_BT}=\sum_{i=1}^{M}\left(n_i\ln\left(\nu_i n_i\right)+\omega_i n_i\right)
   +\sum_{j=2}^M\sum_{i=1}^{j-1}\chi_{ij}n_in_j 
   \label{eq:3}
\end{align}
containing the mixing entropy, the internal free energies $\omega_i$, and the mean-field interactions among the charged components of interaction strength $\chi_{ij}$ \cite{Safran2018, Deviri2021}.

Thermodynamic equilibrium states are characterized by ${\delta F}/{\delta n_i}=0$ and ${\delta F}/{\delta \psi}=0$, leading to
\begin{align}
\mu_i^{\text{el}}&=\frac{\partial f}{\partial n_i} - \kappa_i \nabla^2 n_i + q_ie\psi - \frac{1}{2}\frac{\partial\varepsilon}{\partial n_i} \left(\nabla\psi\right)^2\, , 
\label{eq:4}
\\
\lambda &=\psi \, . 
\label{eq:4b} 
\end{align}
Here, the constant $\mu_i^{\text{el}}$ is the exchange electrochemical potential and the Lagrange multiplier for the Poisson equation is the electrostatic potential $\psi$. When phases coexist, the value of $\mu_i^{\text{el}}$ corresponds to the slope of the Maxwell construction and the osmotic pressures balance between the phases.

To solve the governing equations (i.e., the Poisson equation and the equations provided by constant exchange electrochemical potential conditions in Eq.~\eqref{eq:4}), we use Neumann boundary conditions for the electrostatic potential and the concentrations:
\begin{align}
  \left(\vect{\nabla}\psi\cdot\vect{n}\right)|_{\partial\Omega} &= 0\, , 
  \label{eq:5}
  \\
    \left(\vect{\nabla}n_i\cdot\vect{n}\right)|_{\partial\Omega} &= 0 \, ,
    \label{eq:6}
\end{align}
\end{subequations}
where $\vect{n}$ is the unit outward normal to the system boundary $\partial\Omega$ enclosing the system volume $V$. The Neumann boundary condition for the electrostatic potential ensures that the electric field normal to the boundaries vanishes corresponding to zero surface charge. Integrating the Poisson Eq.~\eqref{eq:1} over the system volume $V$ using the boundary condition~\eqref{eq:5} leads to overall electroneutral systems with $\int d^3x\,\rho(\vect{x})=0$. Note that we do not impose electroneutrality locally, i.e.,  $\rho(\vect{x})\not=0$ in general. The Neumann boundary condition for the concentration field corresponds to an inert boundary that has no affinity to adhere or repel  components.

In summary, Eqs.~\eqref{eq:theory} govern the position-dependent electrothermodynamics of a non-dilute mixture composed of charged components where a dense coacervate phase can coexist with a dilute phase. This theoretical framework extends the classical Poisson-Boltzmann theory \cite{Markovich2021} to non-dilute conditions where interactions among the charged components affect the electrostatic potential.

\section{Results\label{Sec:3}}

\begin{figure*}[!hbt]
   \includegraphics[width=\textwidth]{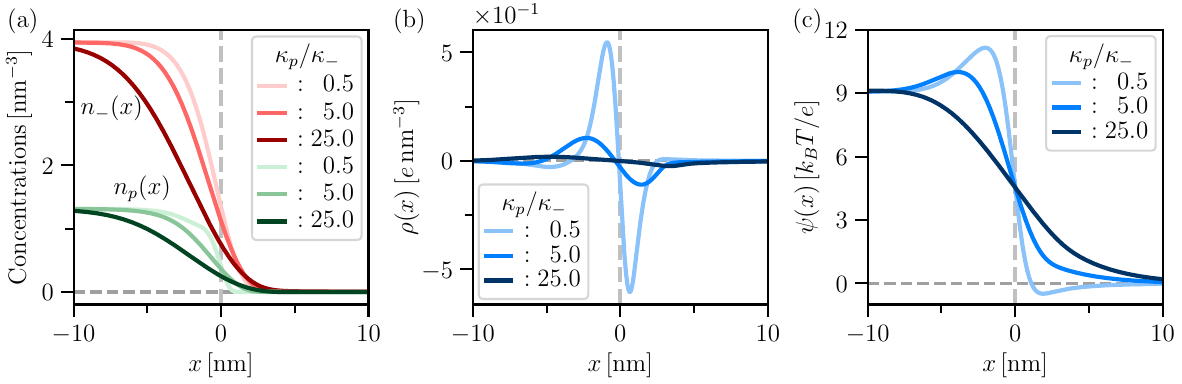}
   \caption{\textbf{Interfacial charge separation of simple coacervates without salt.} 
   \textbf{(a)} Concentration profiles of charged components $n_i(x)$, where ($p$) and ($-$) indicate the macromolecule and counterion, respectively. For increasing $\kappa_p$, the profiles steepen around the interface indicated by a vertical dashed line.
   \textbf{(b)} The charge density $\rho(x)=e\left(q_p n_p(x) + q_-n_-(x)\right)$ indicates separation of charged components in the interfacial domain for small enough $\kappa_p$. Only in this case, the free energy penalty is low enough such that interactions of the charged components with the solvent can give rise to a non-zero charge density. 
   \textbf{(c)} The electrostatic potential $\psi(x)$ has a complex shape within the interfacial domain. There is a well in the negatively charged domain and a barrier in the positively charged domain. The potential is set to zero far in the dilute phase. The interface position is defined where the electrostatic potential is half the value deeply inside the dense coacervate phase.  
 }
   \label{Fig:2}
\end{figure*}

In this work, we use our theoretical framework given by Eqs.~\eqref{eq:theory} to study simple coacervates (Sec.~\ref{Sec:3A}) and complex coacervates (Sec.~\ref{Sec:3B}). For simplicity, we consider flat interfaces and a one-dimensional system with a position $x$. We determine  the position-dependent electrostatic potential $\psi(x)$ together with concentration fields $n_i(x)$ of all the different charged components. In all studies, we use a relative solvent permittivity $\varepsilon_r=80$ corresponding to pure water, and a solvent molecular volume $\nu_s=0.03\,\mathrm{nm}^3$. When salt ions or counter-ions are present in the system, their molecular volumes are for simplicity equal to the solvent. The internal free energies $\omega_i$ do not affect the results as they can be absorbed into the Lagrange multipliers $\mu_i^{\text{el}}$. In our studies on simple coacervates, we choose macromolecules carrying a positive charge. In general, macromolecules such as proteins can be positively or negatively charged under physiological conditions. This diversity arises because, for example, the human cytosol has a $\mathrm{pH}\simeq 7.2$~\cite{Casey2010} and the human proteome shows a bimodal distribution with two major peaks located at $\mathrm{pI}\simeq 6.0$ and $8.25$ \cite{Kurotani2019}.

\subsection{Simple coacervates\label{Sec:3A}}

\subsubsection{Minimal model for simple coacervates \label{Sec:3A1}}

We first discuss a minimal model for a simple coacervate that contains a single type of charged macromolecule ($p$), a neutralizing counterion ($-$), and a solvent component ($s$). Phase separation in such a system takes place for sufficiently large and positive interaction parameters $\chi_{ps}$. For the numerical examples discussed below, we use $\chi_{ps}=1.5\,\mathrm{nm}^{3}$ together with $\chi_{-s}=-0.09\,\mathrm{nm}^{3}$, $\chi_{p-}=-1.0\,\mathrm{nm}^{3}$, and the volume ratios $\nu_p/\nu_s=\nu_p/\nu_-=20$. The counterions are assumed to be monovalent, e.g., $\ce{OH-}$, with $q_-=-1$, and the macromolecule has a charge $q_p=+3$ if not stated otherwise. For such parameters, we numerically solved Eqs.~\eqref{eq:theory}.

Coexistence of a dense coacervate phase and a dilute phase is reflected in the concentration profiles of charged components $n_i(x)$ which are depicted in Fig.~\ref{Fig:2}(a) for different $\kappa_p/\kappa_-$ ratios. Both charged components show a pronounced change highlighting the interface between the dense coacervate phase and the dilute phase. The interface is indicated by a vertical grey dashed line and is defined as the position where the electrostatic potential $\psi(x)$ takes half the value between dense coacervate and dilute phase. The concentration profiles are not symmetric and deviate from the classical interface profile obtained for a symmetric binary Ginzburg-Landau free energy \cite{Bray2002, Weber2019}. The reason is that the mixture is ternary and molecular volumes of the charged components are different. For decreasing relative gradient cost ratio $\kappa_p$ (the ratio $\kappa_p/\kappa_-$ is varied by only varying $\kappa_p$ while keeping $\kappa_-$ fixed), the concentration profiles become steeper around the interface as the energy cost associated with keeping gradients goes down. We also note that although the relative gradient cost ratio $\kappa_p/\kappa_-$ is varied by only varying $\kappa_p$ while keeping $\kappa_-$ fixed, both $n_p$ and $n_-$ show the same behavior with changing $\kappa_p$.

A key finding of our work is that charges can separate within the interfacial domain which is evident in the spatially varying charge density $\rho(x)$ (Fig.~\ref{Fig:2}(b)). While the mixture is charge-neutral ($\rho(x)\simeq 0$) deep within the dense coacervate and dilute phase, we find multiple domains with a positive or negative charge density. The gradient cost $\kappa_p$ essentially determines the shape and the amplitudes of the charge density. For the considered parameters, there is a pronounced negatively charged domain at $x>0$, and a pronounced positively charged domain at $x<0$.

This separation of charge leads to a complex behavior of the electrostatic potential $\psi(x)$ within the interfacial domain (Fig.~\ref{Fig:2}(c)). The negatively charged domain at $x>0$ gives rise to a potential well relative to the reference potential in the dilute phase that is chosen to be zero. On the contrary, the positively charged domain at $x<0$ causes a barrier of the electrostatic potential relaxing toward the Donnan potential $\psi_D$ inside the dense phase \cite{Bagotsky2006, Vis2014}. The electrostatic potential well and barrier implies that a positive test charge would get attracted into the well and repelled by the potential barrier. The existence and strength of both characteristics is set by the cost ration $\kappa_p$. For large values, the well and the potential barrier vanish. The reason is that the charge density in the interface approaches concomitantly to zero (Fig.~\ref{Fig:2}(b)) since the free energy penalty for gradients becomes too large for increasing $\kappa_p$.

The charge separation at the interface can be understood in the following way. Due to differences in the interaction between components and their concentration differences in the two phases, a chemical potential difference arises between the phases. To balance this, an electrostatic potential difference between the two phases created by charge separation at the interface develops such that the resulting electrochemical potential becomes equal everywhere. Although not shown here, all the properties presented in Fig.~\ref{Fig:2} qualitatively remain the same for a system with $\kappa_p/\kappa_-=\nu_p/\nu_-=\left|q_p/q_-\right|=\omega_p/\omega_-=1$, but having slightly different Flory-Huggins interaction parameters. As shown by Aerov \textit{et al.} \cite{Aerov2007-1}, similar phenomenon can happen for interfaces separating ionic and non-ionic liquids.

\subsubsection{Impact of salt on simple coacervates \label{Sec:3A2}}

Now we consider a system that, in addition to the positively charged macromolecules and its counterions, contains salt. For simplicity we consider a monovalent salt and do not make any distinction between the counterions and the anions released by the salt. The resulting quaternary mixture thus contains macromolecules ($p$), cations ($+$), anions ($-$), and solvent ($s$) molecules. For results discussed in this section, we choose the molecular volumes and the interaction parameters as follows: $\nu_p/\nu_s=\nu_p/\nu_+=\nu_p/\nu_-=30$, and $\chi_{ps}=1.9\,\mathrm{nm}^{3}$, $\chi_{p+}=1.8\,\mathrm{nm}^{3}$, $\chi_{p-}=-1.8\,\mathrm{nm}^{3}$, $\chi_{+-}=\chi_{+s}=\chi_{-s}=-0.09\,\mathrm{nm}^{3}$. We also set $\kappa_+=\kappa_-=\kappa$ and unless stated otherwise, use $\kappa=15\,k_BT\mathrm{nm}^5$.

\begin{figure*}[t]
   \includegraphics[width=\textwidth]{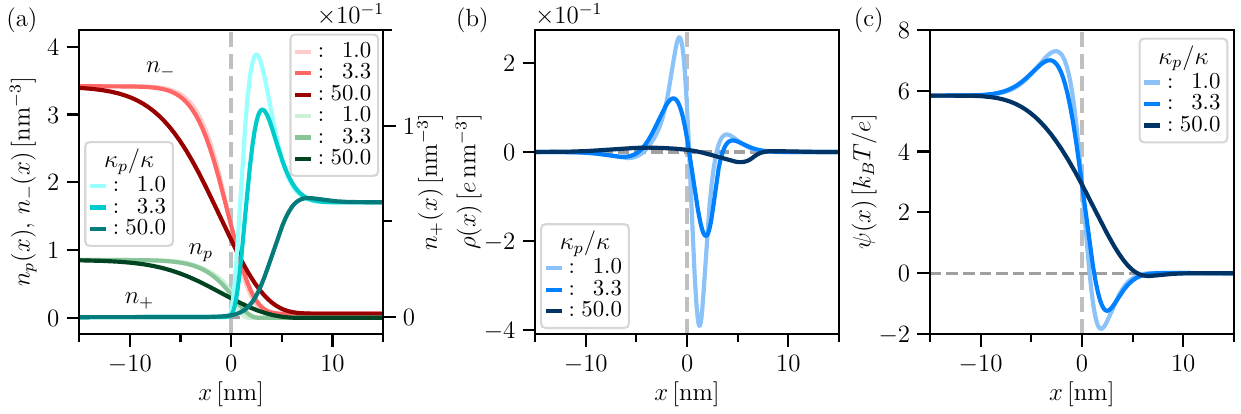}
   \caption{\textbf{Interfacial charge separation of simple coacervates with salt.} 
   \textbf{(a)} In contrast to the case without salt (Fig.~\ref{Fig:2}), the concentration profiles of charged components $n_i(x)$ ($i=p,-,+$) can vary in a non-monotonic fashion within the interfacial domain; the interface is indicated by the vertical dashed line. The dense coacervate phase is enriched in positively charged macromolecules together with its neutralizing counter-ions ($-$), while the positively charged co-ions ($+$) accumulate adjacent to the interface toward the dilute phase. This accumulation enhances for decreasing $\kappa_p$.
   \textbf{(b)} The charge density $\rho(x)=e\left(q_p n_p(x) + q_-n_-(x) + q_+ n_+(x)\right)$ shows, as in the case without salt, that charges separate in the interfacial domain and that charge separation is more pronounced for smaller $\kappa_p$.
   \textbf{(c)} The associated electrostatic potential $\psi(x)$ exhibits  potential wells and barriers that are even more pronounced compared to the case without salt. The plots correspond to a system with salt concentration $c_{\mathrm{salt}}=100\,\mathrm{mM}$, a macromolecular charge $q_p=+4$ and average concentration $\overline{n}_p=0.1\,\mathrm{mM}$, and ion charges $q_{\pm}=\pm 1$.}
   \label{Fig:3}
\end{figure*}

The  concentration profiles of charged components $n_i(x)$ in simple coacervates with salt ($i=p,+,-$) show a complex behavior close to the interface (Fig.~\ref{Fig:3}(a)), where the interface is indicated by a vertical dashed line. The macromolecule concentration $n_p(x)$ decreases monotonically together with the oppositely charged counter-ion ($-$) when passing from the dense coacervate phase toward the dilute phase. This coupled behavior is a result of the attractive electrostatic interactions. Interestingly, the positively charged co-ions ($+$) vary non-monotonically, i.e., ($+$)-ions accumulate right outside the coacervate phase. This accumulation is a result of a macromolecule-poor layer right outside the coacervate and is formed by the attractive electrostatic interactions between oppositely charged ions. This variation vanishes with increasing cost ratio $\kappa_p$ as the profile of macromolecules at the interface flattens and thereby contributes to neutralizing the interfacial domain.

The charge density $\rho(x)$ and the electrostatic potential $\psi(x)$ have a similar qualitative behavior as compared to the case without salt. Multiple charge layers of alternating charge develop within the interfacial region (Fig.~\ref{Fig:3}(b)) and the electrostatic potential can vary non-monotonically with potential wells and barriers (Fig.~\ref{Fig:3}(c)). These profile characteristics vanish for increasing $\kappa_p$ as it quantifies the free energy penalty for profile heterogeneities. However, the presence of salt quantitatively makes charge separation more pronounced, i.e., it is observed for even higher values of $\kappa_p$. 

\begin{figure*}[tb]
   \includegraphics[width=\textwidth]{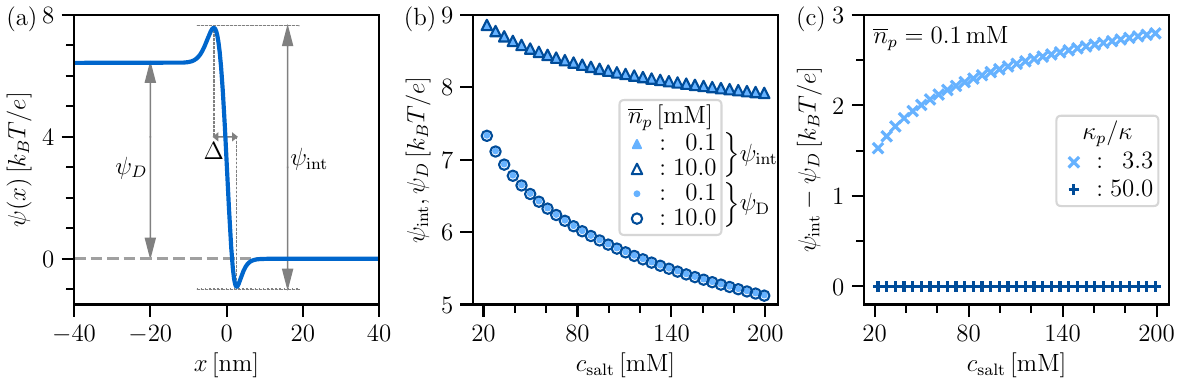}
   \caption{
   \textbf{Impact of salt on the electrostatic potential of simple coacervates.}
   \textbf{(a)} Illustration of the Donnan potential $\psi_D$, the interfacial potential $\psi_{\text{int}}$ and the interfacial width $\Delta$ for a representative electrostatic potential profile $\psi(x)$ obtained for simple coacervates. 
   \textbf{(b)} The Donnan potential $\psi_D$ and the interfacial potential $\psi_{\text{int}}$ decrease with salt concentration $c_{\text{salt}}$ since salt screens the electrostatic interactions. The average macromolecule concentration $\overline{n}_p$ in the system hardly affects $\psi_D$ or $\psi_{\text{int}}$ as it mainly alters the size of the dense coacervate phase.
   \textbf{(c)} For a large value of $\kappa_p=50$, the potential profile varies monotonically and $\psi_{\text{int}}-\psi_D$ vanishes. Thus, there are no additional potential barriers and wells. For small enough values of $\kappa_p$, the difference between Donnan and interfacial potential $\psi_{\text{int}}$ and $\psi_D$ is non-zero and increases with salt concentration $c_\text{salt}$. The reason is that salt reduces the electrostatic potential by screening, which is more pronounced in the bulk phases due to the ions' gradient costs. We fixed average macromolecule concentration $\overline{n}_p=0.1\,\mathrm{mM}$. For all the plots here, macromolecule charge $q_p=4$, co/counter-ion charges $q_{\pm}=\pm 1$ are used.}
   \label{Fig:4}
\end{figure*}

\begin{figure*}[tb]
   \includegraphics[width=\textwidth]{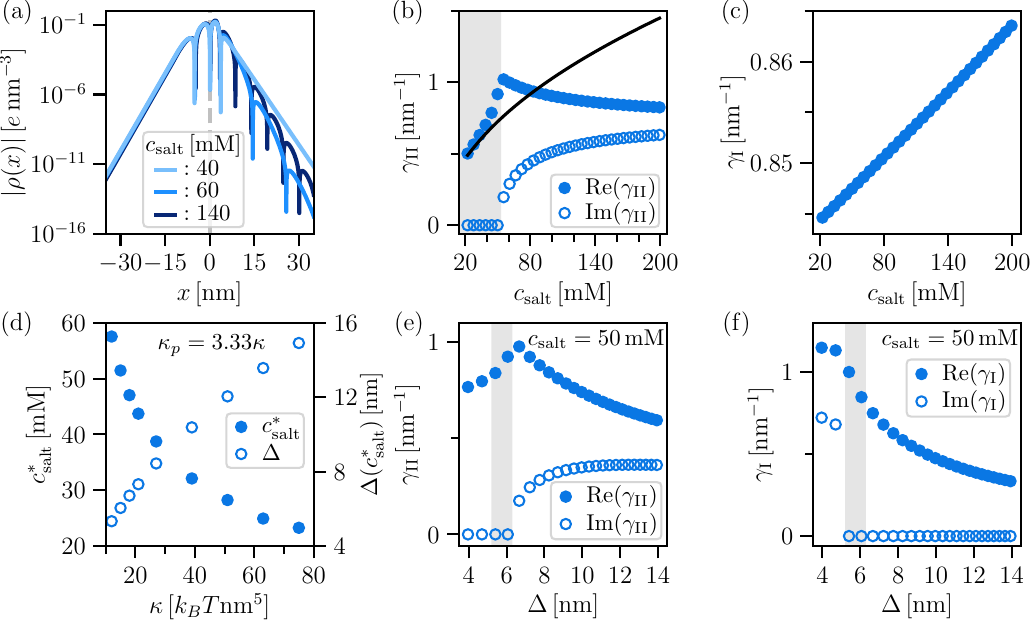}
   \caption{
   \textbf{Layers of alternating charge density in the dense coacervate and dilute phase.}
   \textbf{(a)} Logarithmic representation of the absolute value of the charge density $|\rho(x)|$ around the interfacial domain indicates layers of alternating charge density. The amplitude of charge density $|\rho(x)|$ within layers decay approximately exponentially with characteristic length scales $\Re(\gamma_\text{I/II})^{-1}$ that are phase-dependent. The width of charged layers that can be characterized by $\Im(\gamma_\text{I/II})^{-1}$, decreases with increasing salt concentration $c_{\text{salt}}$.
   \textbf{(b)} In the dilute phase, $\Re(\gamma_\text{II})$ shows a non-monotonic behavior. A non-zero $\Im(\gamma_\text{II})$ indicates the existence of layers of alternating charge. Layers only occur for salt concentrations above the threshold $c^*_{\text{salt}}$. For $c_{\text{salt}}>c^*_{\text{salt}}$, $\Re(\gamma_\text{I/II})$ decreases with salt indicating a deviation from classical Debye-H\"uckel theory. Here we used $\kappa=15\,k_BT\,\mathrm{nm}^5$.
   \textbf{(c)} In the coacervate phase, $\Re(\gamma_\text{I})$ changes only weakly with salt and there are no layers deep within the coacervate phase ($\Im(\gamma_\text{I})=0$).
   \textbf{(d)} The threshold salt concentration $c_{\text{salt}}^*$ decreases with increasing interfacial width $\Delta$, where $\Delta$ can be changed by the gradient cost $\kappa$.
   \textbf{(e,f)} Increasing the interfacial width $\Delta$ makes the system crossing from a situation where alternating charged layers solely exist inside the coacervate, over a regime without any layers, to a regime with layers present exclusively in the dilute phase. For all panels, we use $\overline{n}_p=0.1\,\mathrm{mM}$, $\kappa_+=\kappa_-=\kappa$, $\kappa_p=3.33\kappa$. In panels \textbf{(e)} and \textbf{(f)}, the interfacial width $\Delta$ is varied by varying $\kappa$ in the range $(6-75)\,k_BT\,\mathrm{nm}^5$. Further parameters are the same as in Fig.~\ref{Fig:4}.
   The grey shaded regions in panels \textbf{(b), (e), (f)} indicate situations where the imaginary part of the decay constant is zero in both phases.}
   \label{Fig:5}
\end{figure*}

\begin{figure*}[tb]
   \includegraphics[width=\textwidth]{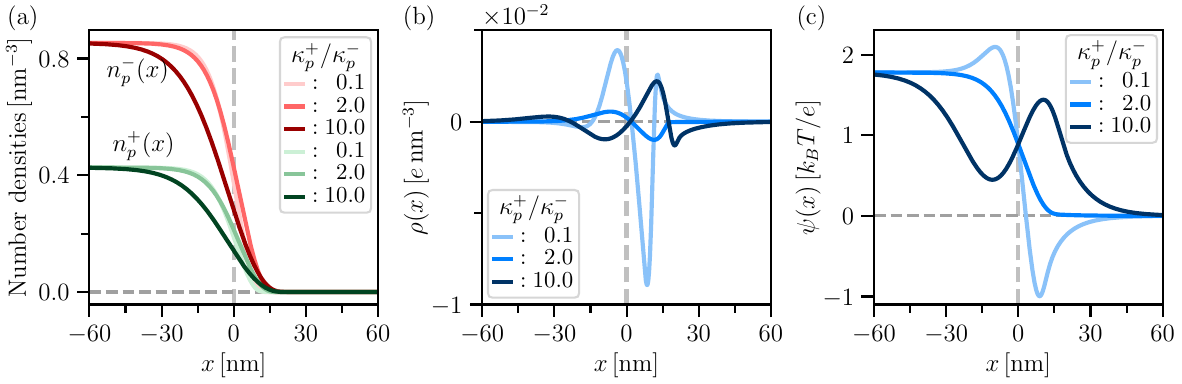}
   \caption{\textbf{Interfacial charge separation of complex coacervates.}
   \textbf{(a)} Monotonic decrease of concentration profiles of the macromolecules $n_p^\pm(x)$ within the interfacial domain. 
   \textbf{(b)} The charge density $\rho(x)=e\left(q_p^+n_p^+(x) + q_p^-n_p^-(x)\right)$ shows complex spatial variation inside the interfacial domain indicating that charge separation can also occur for complex coacervation. By altering the relative gradient cost $\kappa_p^+$ of the oppositely charged macromolecules, the charged layers can swap their sign. 
   \textbf{(c)} This swapping is evident in the corresponding behavior of the electrostatic potential profile $\psi(x)$ when changing $\kappa_p^+$. The ordering of potential barriers and wells swaps. 
   The vertical dashed line in each plot marks the location of the interface. As parameters, we use $q_p^+=4$, $q_p^-=-2$ together with other parameters mentioned in Sec.~\ref{Sec:3B}.}
   \label{Fig:6}
\end{figure*}

The well and barrier of the electrostatic potential is determined by the salt concentration $c_{\text{salt}}$. To quantify this impact, we introduce the interfacial potential $\psi_\text{int}$ as the potential difference between the barrier and the well minimum and contrast it with the Donnan potential $\psi_D$ (Fig.~\ref{Fig:4}(a)). We also propose  a definition for the   interface width of the coacervate $\Delta$ as the distance between the position of potential well and barrier. We find that the Donnan potential $\psi_D$ and the interfacial potential $\psi_{\text{int}}$ decrease with the salt concentration $c_{\text{salt}}$ (Fig.~\ref{Fig:4}(b)). This decrease is due to screening as more salt reduces the electrostatic interactions of the phase-separated macromolecules. For large salt concentrations, this decrease weakens and crosses over to a logarithmic decay for both $\psi_D$ and $\psi_{\text{int}}$, each decaying with a characteristic salt concentration (Fig.~\ref{Fig:4a} in Appendix~\ref{Sec:A-App}). The decrease of $\psi_D$ and $\psi_{\text{int}}$ with salt is however extremely insensitive to the average macromolecule concentration $\overline{n}_p$. The latter predominantly sets the size of the coacervate without altering the composition significantly. For sufficiently large values of the gradient cost $\kappa_p$ the potential profiles become monotonically varying leading to a vanishing difference $(\psi_{\text{int}}-\psi_D)$ (Fig.~\ref{Fig:4}(c)); see Fig.~\ref{Fig:3}(c) for the corresponding electrostatic potential profiles. However, lowering the $\kappa_p$-values leads to barriers and wells of the electrostatic potential and thereby a non-zero differences $(\psi_{\text{int}}-\psi_D)$. Interestingly, though both the Donnan potential $\psi_D$ and the interfacial potential $\psi_{\text{int}}$ decrease with increasing salt, their difference increases (Fig.~\ref{Fig:4}(c)). This trend indicates that the electrostatic potential difference between the dense coacervate phase and the dilute phase is more affected by salt screening than the interfacial potential. This asymmetry between bulk phases and interfacial domains arise from the additional gradient cost compared to the bulk phases for the counter-ions to accumulate in the interfacial domains.

The salt concentration determines the amount and the size of the layers within the interfacial domain. A logarithmic representation of the absolute value of the charge density $|\rho(x)|$ reveals that multiple layers of alternating charge extend from the interfacial domain towards the dense and dilute phase, respectively~(Fig.~\ref{Fig:5}(a)). Increasing the salt concentration increases the number of such layers and decreases layer width. To quantify the layering with salt, we calculated the real and imaginary part of $\gamma_\text{I/II}$ deeply in the dense coacervate and dilute phase by linearizing the profile of each component $i$ around phase equilibrium $n_i^0$ by writing
\begin{align}
    n_i(x)=n_i^0 +c_i \exp(\gamma x);
    \label{eq:7}
\end{align}
for details see Appendix~\ref{Sec:B-App}. We find that amplitudes of charge density $|\rho(x)|$ in the layers decay approximately exponentially with a characteristics length scale $\Re(\gamma_\text{I/II})^{-1}$; an effect reminiscent of classical electrostatic screening as described by the Debye-H\"uckel theory \cite{Israelachvili2011}, where $\Re(\gamma)^{-1}$ would correspond to the Debye screening length. For our simple coacervate in the presence of salt, $\Re(\gamma_\text{II})$ shows a non-monotonic behavior with salt (Fig.~\ref{Fig:5}(b)). For low salt, it increases with increasing salt concentration which is consistent with Debye-H\"uckel theory. However, when passing a threshold salt concentration $c_\text{salt}^*$, this behavior qualitatively changes as $\Re(\gamma_\text{II})$ decreases with increasing salt concentration. The threshold $c_\text{salt}^*$ coincides with the appearance of a non-zero imaginary part $\Im(\gamma_\text{II})$ which indicates the existence of layers of alternating charge also far away from the coacervate interface. In other words, the bulk phases can be layered reminiscent of Coulomb microphase separation in polymer solutions or of block co-polymers undergoing microphase separation \cite{Leibler1980, Semenov1985, Ohta1986, Borue1988, Joanny1990, Leibler1991, Wittmer1993, Rumyantsev2018, Rumyantsev2020, Grzetic2021, Fredrickson2022}, however, noting that the amplitude of charge density decays exponentially when going away from the interface  (Fig.~\ref{Fig:5}(a)). Charged layers, in general, arise due to competing interactions related to different length scales. For our system, these competing length scales are the Bjerrum length $\ell_B$ due to electrostatic interactions and the gradient costs $\kappa_i$ related to interfacial tension (see Appendix~\ref{Sec:B-App} for details). As derived in Appendix~\ref{Sec:B5-App}, for a ternary system comprising of oppositely charged species ($+$,$-$) in a solvent ($s$) in the large $\kappa_{\pm}$ limit, the decay constant $\gamma$ (defined in Eq.~\eqref{eq:7}) reads:
\begin{align}
   \gamma =\pm\frac{1 \pm i}{\sqrt{2}} \left(\frac{e^2}{\varepsilon}\left(\frac{q_+^2}{\kappa_+^2}+\frac{q_-^2}{\kappa_-^2}\right)\right)^{1/4}.
   \label{eq:8}
\end{align}
Clearly, a competition between the Bjerrum length $\ell_B=e^2/\left(4\pi\varepsilon k_BT\right)$ and the length scale associated with the interfacial width gives rise to complex $\gamma$ leading to oscillatory number density as well as electrostatic potential profiles. In the other limit $\kappa_{\pm}\rightarrow 0$, the decay constant is real (see Appendix~\ref{Sec:B-App} for details). While the layering extends to the bulk, the amplitudes of such layers are exponentially damped implying that the manifestation of layering gets negligible deeply in the respective phases. Moreover, for the considered parameters in Fig.~\ref{Fig:5}(a-c), we have found that layers not necessarily exist in both coexisting phases; here we show an example where the dilute phase is layered while the dense coacervate phase is not (Fig.~\ref{Fig:5}(b,c)). Though layers have been reported in theories at charged solid interfaces \cite{Solis1999, Bier2012, Gavish2016}, layering around liquid interface is distinct as the additional interfacial width is coupled to the spatial variations extending towards both bulk phases.

The interfacial width $\Delta$ of the coacervate strongly affects the threshold $c_\text{salt}^*$ of charge layering (Fig.~\ref{Fig:5}(d)). The interfacial width can be changed by increasing  the parameter characterizing the gradient cost $\kappa$, where we chose $\kappa_+=\kappa_-=\kappa$. Thus, for larger $\Delta$, the threshold for layering, $c_\text{salt}^*$, decreases. 

The interfacial width $\Delta$ also controls whether layers of alternating charge occur in the dilute or in the coacervate phase (Fig.~\ref{Fig:5}(e,f)). For low interfacial width $\Delta$ the dense coacervate phase (Fig.~\ref{Fig:5}(f)) exhibits layers, while the dilute phase does not. Increasing $\Delta$ leads to a domain where none of the phases have layers (gray shaded domain in Figs.~\ref{Fig:5}(b,e,f)). For even larger value of the interfacial width $\Delta$, the coacervate is not layered while the dilute phase has layers of alternating charge (Fig.~\ref{Fig:5}(e)). 

\subsection{Complex coacervates\label{Sec:3B}}

Here we discuss a minimal model for complex coacervation and scrutinize our key finding of charge separation at the interface. In contrast to a simple coacervate, complex coacervation is driven by the attractive electrostatic interaction between (at least) two oppositely charged macromolecules. As a result,  both macromolecule types are enriched inside the coacervate phase compared to the coexisting dilute phase. Our minimal model for a complex coacervate accounts for two oppositely charged macromolecules ($p^+$ and $p^-$) suspended in a solvent ($s$). For the following studies, we consider macromolecule-solvent molecular volume ratios $\nu_p^+/\nu_s=\nu_p^-/\nu_s=20$ and Flory-Huggins parameters $\chi_{p^+s}=1.0\,\mathrm{nm}^{3}$, $\chi_{p^-s}=0.6\,\mathrm{nm}^{3}$, and $\chi_{p^+p^-}=-1.0\,\mathrm{nm}^{3}$.

The concentration profiles of the macromolecules $n_p^\pm(x)$ vary monotonously (Fig.~\ref{Fig:6}(a)). We could not find any additional concentration layers as observed for simple coacervates with salt. However, the charge density can exhibit  spatial variation in the interfacial domain indicating that charges can separate also for complex coacervates (Fig.~\ref{Fig:6}(b)). We have varied the relative gradient cost of the positively to the negatively charged macromolecule, $\kappa_p^+/\kappa_p^-$ by varying $\kappa_p^+$. We find that $\kappa_p^+$ can flip the charge of the layers when passing through the interfacial domain. For $\kappa_p^+\ll \kappa_p^-$, the gradients of the concentration field $n_p^+$ can be steeper compared to those of $n_p^-$. Therefore, right inside the coacervate one has $n_p^+ > n_p^-$, leading to a positively charged layer. In the other limit, i.e., for $\kappa_p^+\gg \kappa_p^-$, the gradients of the concentration field $n_p^+$ become less steep compared to those of $n_p^-$. Therefore, one obtains the swapped case with $n_p^- > n_p^+$ and a negatively charged layer right inside the coacervate. This swapping becomes also evident in the behavior of the electrostatic potential $\psi(x)$ as a function of the relative gradient cost $\kappa_p^+$. While for small $\kappa_p^+\ll \kappa_p^-$, the coacervate phase comprises a potential barrier and the dilute phase a potential well, this swaps for large $\kappa_p^+\gg \kappa_p^-$.

We end our discussion by commenting on the relevance of changing the gradient costs $\kappa_p^{\pm}$ and some of the parameter choices in our study. Biological condensates are often formed by complex coacervation of two macromolecules of rather different molecular volume, e.g., proteins with oligonucleotides such as RNA and DNA. Moreover, interaction parameters among macromolecules and solvent are in general different. Since the gradient costs of such macromolecules $\kappa_p^{\pm}$ depend on their molecule volumes and on their interactions, we expect them to vary suggesting the relevance of our results in Fig.~\ref{Fig:6} for biomolecular condensates. We also note that our qualitative results, i.e., interfacial charge separation and associated potential profile formation rely on the asymmetries in Flory-Huggins parameters, and they are expected to be present even for higher macromolecular charges. As to the used model of complex coacervate, in general, free ionic species (e.g., protons) can be present in the solution. However, the resulting quaternary mixture is very much similar to the simple coacervates with added salt case albeit with one ionic species being larger. Therefore, we do not expect the qualitative features of our results to change in that case.

\section{Conclusions\label{Sec:4}}

In summary, we present a theoretical framework to study profiles of concentrations and electrostatic potential at interfaces of phase-separated  solution containing charged components. We apply this framework to simple and complex coacervates. Our approach  extends the  Poisson-Boltzmann theory that describes the spatial distributions of ions adjacent to charged, solid surfaces or charged colloidal particles. In these systems  counter-ions typically follow a Boltzmann distribution and screen the surface charge. Our extension accounts for the interactions among all charged components taking into account phase coexistence. Our work therefore provides a framework for coacervation at salt concentrations where the Poisson-Boltzmann theory fails \cite{Bostroem2001}. Most importantly, in phase-separated systems, interfaces are associated with free energy costs for gradients of charged and uncharged components. These contributions are lacking in the classical Poisson-Boltzmann theory. 

A similar theoretical framework was recently proposed by Zhang \& Wang \cite{Zhang2021}. However, they focus on electrostatic interaction between charged components and consequently, capture interfacial double layer formation for the case of asymmetric complex coacervates. We show that the phenomenon of interfacial charge separation is much more general and can take place in any system with asymmetric interaction between components. We also show that charge separation can take place away from the interface due to a competition between the interfacial length scales and the Bjerrum length.

The key finding of our theory is therefore the general presence of charge separation in the interfacial domain of \textit{any} type of coacervates, be it simple or complex. Beyond a threshold salt concentration, multiple layers of alternating charges can occur around the interface which  extend into the bulk phases. However, the amplitude of the charge contained in each layer decreases exponentially for increasing distance from the interface. We show that salt regulates the number and the width of such layers of alternating charges. Similar layering has also previously been reported, however in homogeneous systems for large salt concentrations. Charge layering in such homogeneous systems results from correlations beyond mean-field as well as within mean-field description~\cite{Kjellander1992, Attard1993, Solis1999, Aerov2007-2, Evans1994, Bier2012, Gavish2016, Smith2016}. Our studies show that layering can occur at mean-field level with moderate salt concentrations due to phase separation where the interfacial width controls the layer patterns.

Charged layers localized around the interface can affect interfacial transport. An interesting case is the stochastic trajectory of single charged probe molecules diffusing across the interface. This molecule will encounter the electric potential profile with barriers and traps which affects the transport kinetics. The resulting charge-specific reflection or trapping of the probe at the interface suggests charge-dependent transport properties at the interface. In the context of biomolecular condensates forming via phase separation in cells \cite{Hyman2014}, this phenomenon could be used by cells to regulate molecular transport, a property usually associated with membranes~\cite{Nikaido1992, Geise2014}.

Our finding of complex-shaped charged layers at the coacervate interfaces implies that a coacervate is not a colloid with a surface charge surrounded by a (dilute) layer of screening counter-ions. Counter-ions at coacervate interfaces are, in general, non-dilute and participate in the phase separation equally as the macromolecules. In other words, the surface charge is not solely a property of the coacervate, as is typically the case for colloids. For coacervates, the surface charge depends on the interactions and concentrations of all charged components in the system. Thus, changes in salt or  gradient costs alter the coacervate's surface charge and the screening counter-ion distribution. This complexity poses an exciting challenge in calculating the electrophoretic mobility when coacervates are subject to external electric fields \cite{Banerjee2017, Welsh2022, Haren2024}.

Our theoretical framework also paves the way to investigating even more complex phase separation phenomena in non-dilute mixtures composed of charged components. It remains elusive why coacervates hardly undergo coarsening by Ostwald ripening \cite{Abbas2021, Nakashima2021, Chen2023}, and why coacervates can repel each other or stick without fusing \cite{Tena2021}. Moreover, the surface charge of coacervates could also be regulated by pH \cite{Arana2020, Majee2020, Celora2023}, and coacervates can also act as chemical reactors \cite{Tang2013, Ghosh2021, Kubota2022}. Recently, coacervates maintained away from equilibrium were shown to deform to complex shapes, including the formation of liquid shells \cite{Bauermann2023, Bergmann2023}.

Finally, we note that, for simplicity, we have combined a Flory-Huggins model with long range electrostatic interactions to discuss profiles of interfacial charges. However, other approaches have been proposed that also account for an electrostatic free energy beyond mean-field as well as chain correlations along polymer~\cite{Lifshitz1969, Lifshitz1978, deGennes1980, Johner1991, Zhang2021}. It will be interesting to include such effects in our analysis, and additionally account for charge regulation of macomolecules~\cite{Arana2020, Majee2020} to investigate how the charge oscillations at and away from the coacervate interface are affected.

\acknowledgments{
We thank L.\ Hubatsch for fruitful discussion on the topics of interfacial transport, P.\ M.\ McCall for general discussions on electrostatics of protein droplets, M.\ Bier and S.\ Dietrich for discussions on the oscillatory behavior of charge and density profiles. F.\ J\"ulicher acknowledges funding by the Volkswagen Foundation. C.\ Weber acknowledges the European Research Council (ERC) for financial support under the European Union’s Horizon 2020 research and innovation programme (``Fuelled Life'' with Grant agreement No.\ 949021).}

\appendix

\section{$\psi_D$ and $\psi_{\mathrm{int}}$ as function of salt concentration\label{Sec:A-App}}

\begin{figure}[tb]
   \includegraphics[width=7.0cm]{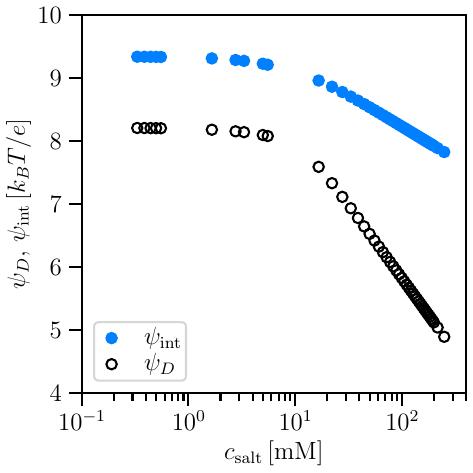}
   \caption{Variation of both the Donnan potential $\psi_D$ and the interfacial potential $\psi_{\text{int}}$ (defined in Fig.~\ref{Fig:4}(a)) as functions of the salt concentration $c_{\text{salt}}$ showing that $\psi_D$ and $\psi_{\text{int}}$ vary logarithmically for large $c_{\text{salt}}$ but with different rates.}
   \label{Fig:4a}
\end{figure}

In Fig.~\ref{Fig:4} of the main text, we have shown the variations of the Donnan potential $\psi_D$ and the interfacial potential $\psi_{\text{int}}$ as functions of the salt concentration $c_{\text{salt}}$ using linear scales. In order to gain further insight into the actual variation, we plot them for an even larger salt concentration range using semi-log scale in Fig.~\ref{Fig:4a}. As the plots suggest, both $\psi_D$ and $\psi_{\text{int}}$ vary logarithmically with $c_{\text{salt}}$ but with different rates.

\section{Decaying charge oscillations towards bulk phases} \label{Sec:B-App}

\subsection{Governing equations \label{Sec:B1-App}}

Let us consider a ternary system consisting of a positively charged species ($+$), a negatively charged species ($-$) and a solvent ($s$). Conservation of volume fractions implies that all of the three volume fractions are not independent of each other. As always, we consider the one for the solvent to be the dependent one which is given by $\left(1-\nu_+ n_+ - \nu_- n_-\right)$. The other two volume fractions or equivalently, the corresponding concentration profiles (as we consider constant molecular volumes $\nu_{\pm}$) together with the electrostatic potential everywhere in the system can be obtained by solving the following equations:
\begin{align}
\nabla^2\psi&=-\frac{\rho}{\varepsilon},\label{eq:B1}\\
\mu_{\pm}^{\text{el}}&=\text{constant}.\label{eq:B2}
\end{align}

\subsection{Exponential ansatz and small deviation assumption \label{Sec:B2-App}}

Away from the interface, all the quantities decay exponentially as
\begin{align}
n_+&=n_+^0+c_+\exp{(\gamma x)} + \mathrm{c.\,c.},\label{eq:B3}\\
n_-&=n_-^0+c_-\exp{(\gamma x)} + \mathrm{c.\,c.},\label{eq:B4}\\
\psi&=\psi^0+c_{\psi}\exp{(\gamma x)} + \mathrm{c.\,c.},\label{eq:B5}    
\end{align}
to their respective bulk values, i.e., to $n_+^0$, $n_-^0$, and $\psi^0$ with a decay constant $\gamma$. Here, $\mathrm{c.\,c.}$ refers to the complex conjugate. As there are three equations to solve for four unknowns ($c_+$, $c_-$, $c_{\psi}$, and $\gamma$), one can obtain only the ratios $\displaystyle{c_p={c_+}/{c_{\psi}}}$ and $\displaystyle{c_m={c_-}/{c_{\psi}}}$. Inserting the exponential profiles (Eqs.~\eqref{eq:B3}--\eqref{eq:B5}) back into the Poisson equation \eqref{eq:B1}, one obtains the condition
\begin{align}
    \gamma^2=-\frac{e\left(q_+c_p + q_-c_m\right)}{\varepsilon}.
    \label{eq:B6} 
\end{align}

\begin{figure*}[tb]
   \includegraphics[width=\textwidth]{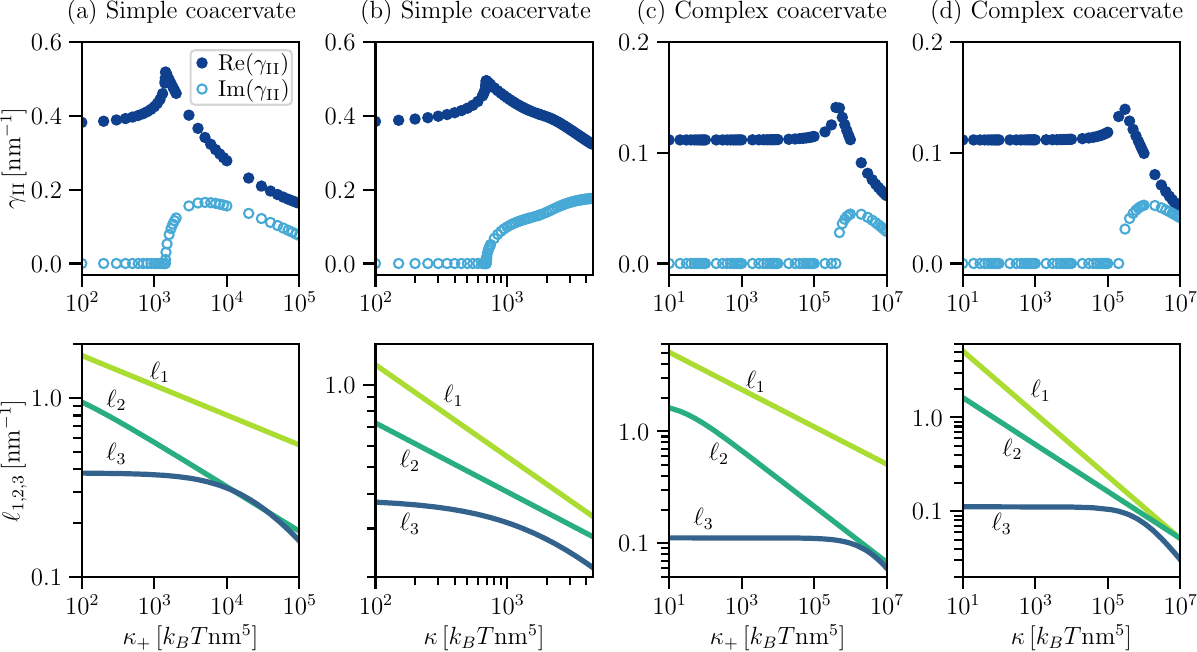}
   \caption{\textbf{Change in length scales corresponding to charge oscillations in the dilute phase.}
   \textbf{Upper panels:} The real and imaginary parts of the characteristic length scales $\gamma_\text{II}$ in the dilute phase as functions of the surface parameter(s) for simple (panels (a) and (b)) and complex coacervates (panels (c) and (d)). In panels (a) and (c), only the surface parameter $\kappa_+$ is varied keeping $\kappa_-=10\,k_BT\,\mathrm{nm}^5$ fixed. In contrast, panels (b) and (d) show the behavior of $\gamma_\text{II}$ when both the surface parameters $\kappa_+=\kappa_-=\kappa$ are varied. In both cases, one sees that the decay constant turns complex from real with increasing surface parameters, or equivalently, with increasing interfacial width.
   \textbf{Lower panels:} Log-log plots showing the variations of the length scales $\ell_1$, $\ell_2$, and $\ell_3$ defined in Eqs.~\eqref{eq:B12} corresponding to the situations considered in the respective upper panel. As one can identify, in each case, the length scale $\ell_3$ changes slope around the region where the length scale $\gamma$ becomes imaginary. The other two length scales, i.e., $\ell_1$ and $\ell_2$ do not show such changes in their slopes.
   Unless stated explicitly, all the parameters correspond to those used in Fig.~\ref{Fig:2} and Fig.~\ref{Fig:6} of the main text for simple and complex coacervates, respectively.}
   \label{Fig:8}
\end{figure*}

As $c_+$, $c_-$, and $c_{\psi}$ are not restricted to be real-valued, both $c_p$ and $c_m$, and consequently, $\gamma$ can also be complex-valued. Please note that a complex $\gamma$ corresponds to oscillatory profiles for the number densities and the electrostatic potential. Next, the exponential profiles are inserted into the equal exchange electrochemical potential conditions (Eq.~\eqref{eq:B2}). The log terms are expanded up to linear order with respect to small deviations from the bulk densities (i.e., in $\exp\left(\gamma x\right)$) in the following way:
\begin{align*}
    \log\left(\nu_in_i\right)&=\log\left(\nu_i n_i^0+\nu_i c_i\exp{(\gamma x)}\right)\notag\\
    &=\log\left[\nu_i n_i^0\left(1+\frac{c_i}{n_i^0}\exp{(\gamma x)}\right)\right]\notag\\
    &=\log\left(\nu_i n_i^0\right)+\log\left(1+\frac{c_i}{n_i^0}\exp{(\gamma x)}\right)\notag\\
    &=\log\left(\nu_i n_i^0\right) + \frac{c_i}{n_i^0}\exp{(\gamma x)} + ...,
\end{align*}
and
\begin{align*}
    &\log\left(1-\nu_+n_+-\nu_-n_-\right)\notag\\
    &=\log\left[1 - \nu_+ n_+^0 - \nu_-n_-^0 - \left( \nu_+c_+ - \nu_-c_-\right) \exp{(\gamma x)} \right]\notag\\
    &=\log\left[\phi_s^0\left(1 - \frac{\nu_+c_+}{\phi_s^0}\exp{(\gamma x)} - \frac{\nu_-c_-}{\phi_s^0}\exp{(\gamma x)}\right)\right]\notag\\
    &=\log\left(\phi_s^0\right)+\log\left(1 - \frac{\nu_+c_+}{\phi_s^0}\exp{(\gamma x)} - \frac{\nu_-c_-}{\phi_s^0}\exp{(\gamma x)}\right)\notag\\
    &= \log\left(\phi_s^0\right) - \frac{\nu_+c_+}{\phi_s^0}\exp{(\gamma x)} - \frac{\nu_-c_-}{\phi_s^0}\exp{(\gamma x)} + ...,
\end{align*}
where $\phi_s^0=1-\nu_+n_+^0-\nu_-n_-^0$. Using these expressions in the constant exchange electrochemical potential conditions and equating the coefficients of $\exp{\left(\gamma x\right)}$ to zero, one obtains: 
\begin{align}
    \kappa_+\gamma^2c_p =& \left[\frac{c_p}{n_+^0}+\frac{\nu_+}{\nu_s}\left(\frac{\nu_+c_p}{\phi_s^0}+\frac{\nu_-c_m}{\phi_s^0}\right)\right]k_BT + \chi_{+-}c_m\notag\\
    &-\chi_{+s}\left(2\frac{\nu_+}{\nu_s}c_p+\frac{\nu_-}{\nu_s}c_m\right)-\chi_{-s}\frac{\nu_+}{\nu_s}c_m + eq_+,
    \label{eq:B7}
\end{align}
and
\begin{align}
    \kappa_-\gamma^2c_m =& \left[\frac{c_m}{n_-^0}+\frac{\nu_-}{\nu_s}\left(\frac{\nu_+c_p}{\phi_s^0}+\frac{\nu_-c_m}{\phi_s^0}\right)\right]k_BT+\chi_{+-}c_p\notag\\
    &-\chi_{-s}\left(2\frac{\nu_-}{\nu_s}c_m+\frac{\nu_+}{\nu_s}c_p\right)-\chi_{+s}\frac{\nu_-}{\nu_s}c_p + eq_-.
    \label{eq:B8}
\end{align}
Equations~\eqref{eq:B6}, \eqref{eq:B7}, and \eqref{eq:B8} form a set of three equations to be solved for three unknowns $c_p$, $c_m$, and $\gamma$. We first solve Eqs.~\eqref{eq:B7}, and \eqref{eq:B8}, to obtain $c_p$ and $c_m$ in terms of $\gamma$ and this is then used in Eq.~\eqref{eq:B6} to obtain the following polynomial equation satisfied by the decay constant $\gamma$:
\begin{align}
    &\varepsilon\kappa_+\kappa_-\gamma^6 - \varepsilon\left(a\kappa_-+d\kappa_+\right)\gamma^4 \notag\\
    &+\left(\varepsilon a d - \varepsilon b^2 + e^2q_+^2 \kappa_- + e^2q_-^2 \kappa_+\right)\gamma^2 \notag\\
    &+e^2\left(2bq_+q_- - q_+^2 d - q_-^2 a\right) =0,
    \label{eq:B9}
\end{align}
where
\begin{subequations}\label{eq:B10}
\begin{align}
    a&=\left(\frac{1}{n_+^0} + \frac{\nu_+^2}{\nu_s\phi_s^0}\right)k_BT - 2\chi_{+s}\frac{\nu_+}{\nu_s},\label{eq:B10:a}\\
    b&=\frac{\nu_+\nu_-}{\nu_s\phi_s^0}k_BT+\chi_{+-} - \chi_{+s}\frac{\nu_-}{\nu_s} - \chi_{-s}\frac{\nu_+}{\nu_s},\label{eq:B10:b}\\
    d&=\left(\frac{1}{n_-^0} + \frac{\nu_-^2}{\nu_s\phi_s^0}\right)k_BT - 2\chi_{-s}\frac{\nu_-}{\nu_s}.\label{eq:B10:c}
\end{align}
\end{subequations}
Equation \eqref{eq:B9} is a polynomial equation of sixth order in $\gamma$ and can have real as well as complex roots in accordance with what one can already infer from Eq.~\eqref{eq:B6}. A similar calculation can be done for a quaternary system which leads to a polynomial equation of higher degree.

\subsection{Complex decay constant $\gamma$ \label{Sec:B3-App}}
In order to get an idea of the nature and origin of the different roots of $\gamma$, we rewrite Eq.~\eqref{eq:B9} in the following way:
\begin{align}
    \left(\frac{\gamma}{\ell_1}\right)^6 - \left(\frac{\gamma}{\ell_2}\right)^4 + \left(\frac{\gamma}{\ell_3}\right)^2 + r = 0
    \label{eq:B11}
\end{align}
with
\begin{subequations}\label{eq:B12}
\begin{align}
    \ell_1^6&=\frac{e^2\left(q_+^2 d + q_-^2 a - 2bq_+q_-\right)|_{\chi\rightarrow 0}}{\varepsilon\kappa_+\kappa_-},\label{eq:B12:a}\\
    \ell_2^4&=\frac{e^2\left(q_+^2 d + q_-^2 a - 2bq_+q_-\right)|_{\chi\rightarrow 0}}{\varepsilon\left(a\kappa_-+d\kappa_+\right)},\label{eq:B12:b}\\
    \ell_3^2&=\frac{e^2\left(q_+^2 d + q_-^2 a - 2bq_+q_-\right)|_{\chi\rightarrow 0}}{\varepsilon a d - \varepsilon b^2 + e^2q_+^2 \kappa_- + e^2q_-^2 \kappa_+},\label{eq:B12:c}\\
    r&=\frac{2bq_+q_- - q_+^2 d - q_-^2 a}{\left(q_+^2 d + q_-^2 a - 2bq_+q_-\right)|_{\chi\rightarrow 0}}.\label{eq:B16:d}
\end{align}
\end{subequations}
The $\chi\rightarrow 0$ condition implies that the expression is calculated in the limit of vanishing Flory-Huggins interaction between all components. The resulting limiting expression can be shown to always have a nonzero and strictly positive value. By construction, $\ell_1$, $\ell_2$, and $\ell_3$ are three length scales associated with the roots of $\gamma$ in Eq.~\eqref{eq:B9}.

\begin{figure*}[tb]
   \includegraphics[width=\textwidth]{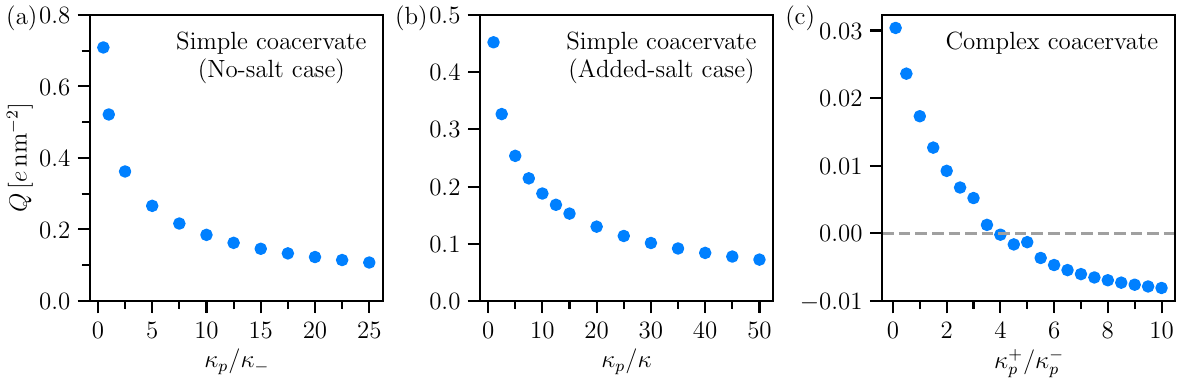}
   \caption{Variation of both the total charge contained within the coacervate phase as function of the relative gradient cost parameters for \textbf{(a)} simple coacervates without salt, \textbf{(b)} simple coacervates with salt as well as for \textbf{(c)} complex coacervates. Whereas for the simple coacervates the total charge of the coacervate remains of the same sign with changing relative gradient cost, for complex coacervates it can even change sign.}
   \label{Fig:9}
\end{figure*}

Figure~\ref{Fig:8} shows the variations of these length scales and the corresponding variation of the characteristic length scale $\gamma_{\mathrm{II}}$ in the dilute phase as function of the gradient cost for a ternary system (either a simple coacervate without salt (panels (a) and (b)) or a complex coacervate (panels (c) and (d))). For both systems, either only one of the gradient costs $\kappa_+$ or both of them, i.e., $\kappa_+=\kappa_-=\kappa$ are varied to change the interfacial width. We note that increasing the gradient cost increases the interfacial width. As one can see from the upper panels of all the plots, with increasing interfacial width, the decay constant $\gamma_{\mathrm{II}}$ becomes complex from being purely real. Quite interestingly, the corresponding lower panels suggest that the length scale $\ell_3$ also changes its slope around the position where $\gamma_{\mathrm{II}}$ changes its behavior. The other two length scales $\ell_{1,2}$ do not manifest such a change. Therefore, the origin of the complex decay constant can be attributed to the length scale $\ell_3$ which, as Eq.~\eqref{eq:B12:c} suggests, arises due to a competition of the electrostatic length scale and of the interfacial width. We note that in the limit of vanishing Flory-Huggins interactions $\chi_{ij}\rightarrow 0$, vanishing molecular volumes $\nu_{\pm}\rightarrow 0$, and vanishing gradient costs $\kappa_{\pm}\rightarrow 0$, the length scale $\ell_3=\sqrt{\left(q_+^2e^2n_+^0 + q_-^2e^2n_-^0\right)/\left(\varepsilon k_BT\right)}$ matches the well-known inverse Debye length.

\subsection{Criterion for obtaining complex decay rate $\gamma$ \label{Sec:B4-App}}
By using a quadratic variable substitution, Eq.~\eqref{eq:B9} can be rewritten as a cubic equation which has complex roots for negative discriminant, i.e., for
\begin{align}
    \frac{1}{\ell_2^8\ell_3^4}+\frac{4r}{\ell_2^{12}}-\frac{4}{\ell_1^6\ell_3^6}-\frac{18r}{\ell_1^6\ell_2^4\ell_3^2}-\frac{27r^2}{\ell_1^{12}}<0.
    \label{eq:B13}
\end{align}
Consequently, $\gamma$ can also have complex (with non-zero real part) values only when the condition in \eqref{eq:B13} is satisfied.

\subsection{The limits of large and small gradient costs \label{Sec:B5-App}}
In the limits of large and small gradient costs, the behavior of the decay constant $\gamma$ can also be analytically understood.
\subsubsection{Large gradient costs}
In the limit of large gradient costs $\kappa_{\pm}$, we make the following reasonable ansatz:
\begin{align}
    \gamma\sim\kappa^{\alpha},~~~\alpha<0.
\end{align}
Using this in Eq.~\eqref{eq:B9}, one obtains
\begin{align}
    &\mathcal{O}\left(\kappa^{2+6\alpha}\right) + \mathcal{O}\left(\kappa^{1+4\alpha}\right) + \mathcal{O}\left(\kappa^{2\alpha}\right) \notag\\
    &+ \mathcal{O}\left(\kappa^{1+2\alpha}\right) + \mathcal{O}\left(\kappa^{0}\right) = 0
    \label{eq:B14}
\end{align}
In order to get meaningful solutions for $\gamma$ only one term in this expression cannot dominate. Therefore, at least two of the terms scale equally in Eq.~\eqref{eq:B14} and the remaining ones grow slower. Keeping in mind that $2\alpha<(1+2\alpha)$ as well as $(1+4\alpha)<(1+2\alpha)$ for $\alpha<0$, one can have the following three situations.
\paragraph*{\emph{\ul{Case I: $\kappa^{2+6\alpha}=\kappa^{1+2\alpha}$}:}}
In this case $\alpha=-\frac{1}{4}$ and the polynomical equation \eqref{eq:B9} simplifies to
\begin{align*}
    \varepsilon\kappa_+\kappa_-\gamma^6 + e^2\left(q_+^2\kappa_- + q_-^2\kappa_+\right) \simeq 0.
\end{align*}
Its solution 
\begin{align}
    \gamma&\simeq\pm\frac{1\pm i}{\sqrt{2}}\left(\frac{e^2}{\varepsilon}\left(\frac{q_+^2}{\kappa_+} + \frac{q_-^2}{\kappa_-}\right)\right)^{1/4}\notag \\
    &\simeq\pm\frac{1\pm i}{\sqrt{2}}\left(4\pi \ell_Bk_BT\left(\frac{q_+^2}{\kappa_+} + \frac{q_-^2}{\kappa_-}\right)\right)^{1/4}
    \label{eq:B15}
\end{align}
is discussed in the main text in detail; see Eq.~\eqref{eq:8}. Essentially, this solution is complex implying an oscillatory decay of the quantities in Eqs.~\eqref{eq:B3} -- \eqref{eq:B5} as well as of the corresponding charge density and is due to a competition of the electrostatic interaction and interfacial tension.
\paragraph*{\emph{\ul{Case II: $\kappa^{2+6\alpha}=\kappa^{0}$}:}}
This condition implies $\alpha=-\frac{1}{3}$. However, plugging this value of $\alpha$ in the different terms, one obtains 
\begin{align*}
    \kappa^{2+6\alpha}=\kappa^0<\kappa^{1/3}=\kappa^{1+2\alpha}
\end{align*}
implying that the third term dominates over the other two. Thus, this case cannot take place.
\paragraph*{\emph{\ul{Case III: $\kappa^{1+2\alpha}=\kappa^{0}$}:}}
Solving this, one obtains $\alpha=-\frac{1}{2}$ and plugging this back into Eq.~\eqref{eq:B9} leads to the simplified form
\begin{align*}
    e^2\left(q_+^2\kappa_- + q_-^2\kappa_+\right)\gamma^2 + e^2\left(2bq_+q_- - q_+^2d - q_-^2a\right) \simeq 0.
\end{align*}
The solution $\gamma\simeq\pm\sqrt{\frac{q_+^2d + q_-^2a - 2bq_+q_-}{q_+^2\kappa_- + q_-^2\kappa_+}}$ of this quadratic equation can never be complex. It is either real or purely imaginary and stems from a competition of short-range Flory-Huggins interactions and the interfacial tension.

\subsubsection{Small gradient costs}
For small gradient costs, i.e., for $\kappa_{\pm}\rightarrow 0$, Eq.~\eqref{eq:B9} simplifies to
\begin{align*}
    \varepsilon\left(ad-b^2\right)\gamma^2 + e^2\left(2bq_+q_- - q_+^2d - q_-^2a\right) \simeq 0,
\end{align*}
and the resulting decay rate $\gamma\simeq\pm\sqrt{\frac{e^2\left(q_+^2d + q_-^2a - 2bq_+q_-\right)}{\varepsilon\left(ad-b^2\right)}}$ originating due to a competition of the electrostatics and short-range Flory-Huggins interactions is either real or purely imaginary. Therefore, one cannot obtain oscillatory profiles in the sharp-interface limit.

\section{Total charge $Q$ inside the coacervate phase \label{Sec:C-App}}

In Fig.~\ref{Fig:9}, we plot the total charge $Q = \int_{-\infty}^0 dx \rho(x)$ contained inside the coacervate phase for three different cases (simple coacervates with or without salt and complex coacervates). As one can see, in all three cases, the total charge $Q$ decreases with increasing gradient cost ratios. In addition, it can even flip sign for complex coacervates. The kinks in the curve corresponding to complex coacervates (Fig.~\ref{Fig:9}(c)) correspond to changes in the shape of the electrostatic potential profiles as depicted in Fig.~\ref{Fig:6}(c). We note that the position of the interface, i.e., the location $x=0$ is defined based on the electrostatic potential profiles. In case the potential profile encounters the value $\psi_D/2$ more than once, we choose the middle one to be the location of the interface.

\bibliography{paper}

\end{document}